\documentclass[a4paper,twocolumn,showpacs,amsmath,pra,aps,10pt,amssymb,nofootinbib]{revtex4-1} 
\usepackage[T1]{fontenc}
\usepackage[latin9]{inputenc} 
\usepackage{times}
\usepackage{color}
\usepackage{xspace}
\usepackage{amssymb,amsmath}
\usepackage{amsbsy}
\usepackage[pdftex]{graphicx}
\usepackage{float}
\usepackage{verbatim}

\newcommand{\Cdd}{C_{\mathrm{dd}}}
\newcommand{\br}{\mathbf{r}}
\newcommand{\bx}{\mathbf{x}}
\newcommand{\bk}{\mathbf{k}}

\newcommand{\vt}[1]{U_{\mathrm{tr}}(#1)}
\newcommand{\ldb}{\lambda_{\mathrm{dB}}}
\newcommand{\half}{\frac{1}{2}}

\newcommand{\kt}{k_BT}
\newcommand{\aho}{\ensuremath{a_{\mathrm{ho}}}\xspace}

\newcommand{\UD}{U_{\mathrm{dd}}}
\newcommand{\UDt}{\tilde{U}_{\mathrm{dd}}}

\newcommand{\evar}{V_\mathrm{eff}\xspace}

\newcommand{\bosee}[3][{}]{\ensuremath{\zeta^{#1}_{#2}\negthinspace\left(e^{#3}\right)}\xspace}
\begin{document}
\title{A local exchange theory for trapped dipolar gases}
\author{D.~Baillie} 
\author{P.~B.~Blakie}  

\affiliation{Jack Dodd Centre for Quantum Technology, Department of Physics, University of Otago, Dunedin, New Zealand.}

\begin{abstract}
We develop a practical Hartree-Fock theory for trapped Bose and Fermi gases that interact with dipole-dipole interactions. This theory is applicable at zero and finite temperature. Our approach is based on the introduction of  local momentum distortion fields that characterize the exchange effects in terms of a local effective potential. We validate our theory against existing theories, finding excellent agreement with full Hartree-Fock calculations.
\end{abstract}


\maketitle

 
 {\bf Introduction:}  
Phenomenal progress in the production of ultra-cold quantum gases with magnetic \cite{Griesmaier2005a,*Bismut2010a,*Pasquiou2011a,Mingwu2011a,Aikawa2012a,Lu2012a}  and electric \cite{Aikawa2010a,Ni2008a} dipoles has opened up an important new manybody system \cite{Lahaye2007a}.  The key feature of these gases is that the constituent particles interact via a dipole-dipole interaction (DDI) that is long-ranged and anisotropic.

There has been considerable success in the development of theory for dipolar Bose-Einstein condensates, in which all the atoms occupy a single mode that is  described by the meanfield Gross-Pitaevskii equation \cite{Goral2000a,*Kawaguchi2006a,*Ronen2006b,*Kawaguchi2006b,*Yi2006a,*Kawaguchi2007a,*Wilson2009a,*Parker2009a,*ODell2004a}. However in situations where many modes are occupied (i.e.~a Bose gas at finite temperature or a Fermi gas)  the meanfield treatment of the non-local exchange interaction is technically challenging. This issue is most pronounced in the experimentally relevant case of trapped samples where both direct and exchange effects contribute. To date, calculations including exchange have been performed by two groups for normal Bose and Fermi gases within the Hartree-Fock (HF) approximation \cite{Zhang2009a,Zhang2010a,Baillie2010b,Baillie2012a}, and for small quasi-two dimensional condensates \cite{Ticknor2012a} within the HF-Bogoliubov-Popov approach.  These calculations are numerically intensive and are only practical in cases where the dimensionality is reduced, either through cylindrical symmetry or by tight confinement.  
Some simple variational   \cite{Miyakawa2008a,*Endo2010a,*Lima2010a} and phenomenological  \cite{Zinner2011a} treatments of exchange have been investigated (see comparisons to HF calculations in \cite{Zhang2009a,Zhang2010a,Baillie2010b,Baillie2012a}).   We also note the application of beyond-meanfield Monte-Carlo methods  to two-dimensional gases \cite{Filinov2010a,*Cinti2010a,*Henkel2012a}.

Exchange effects are predicted to cause dipolar gases to undergo momentum space magnetostriction \cite{Baillie2012a}, and have a significant role in mechanical stability \cite{Zhang2009a,Zhang2010a}.
A number of studies of homogeneous Fermi systems have shown the importance of exchange for various phase transitions and other phenomena  \cite{Babadi2011a,*Baranov2011a,*Zinner2012a,*Parish2012a,*Chan2010a,*Kestner2010a,*Ronen2010a,*Cheng2010a,*Shi2010a,*Liao2010a}, but extending these predictions to the trapped system remains an outstanding problem.

Here we report on the development of a tractable Hartree Local-Fock (HLF)  theory for trapped dipolar gases that  accurately describes both direct and exchange interactions. Our theory is based on the semiclassical HF approximation (avoiding the need to diagonalize for modes), a  theory that has been extensively applied to gases with contact interactions \cite{Giorgini1996a,*Giorgini1997a,*Giorgini1997b,*Dalfovo1997a,*Giorgini2008a}, and provides a good description of experiments (e.g.~see \cite{Gerbier2004,*Gerbier2004b}). The HLF theory is derived by introducing a pair momentum distortion fields that simplify the exchange term to a local potential.  This approach provides insight into the manifestation of exchange interactions in the dipolar gas, and opens a path for developing meanfield theories in the superfluid regime. 

We validate the HLF theory against HF and Hartree calculations for Bose and Fermi systems at zero and finite temperature. 
The HLF theory is vastly faster and more resource efficient: a HF calculation taking 40 hours is reduced to $2$ seconds with HLF.\footnote{Assuming a cylindrically symmetric trap to make HF calculations feasible, with $(N_\rho, N_z)$ grid points in (radial, axial) directions (both momentum and spatial directions for HF and just spatial directions for HLF), for HF calculations the slow step is calculating $\Phi_E(\bx,\bk)$ which is $O[(N_\rho N_z)^3]$, and for HLF the slow step is calculating $\delta(\bx)$ which is $O(N_\rho N_z)$. The example times given are for fixed $\mu$ with $N_\rho=N_z=100$.} 

 {\bf System:} 
We consider a gas of spin polarized particles that interact by a DDI of the form
\begin{align}
\UD(\br)=\frac{\Cdd}{4\pi}\frac{1-3\cos^2\theta}{|\br|^3},\label{e:udd}
\end{align}
where $\Cdd=\mu_0\mu^2_m$ for magnetic dipoles of strength $\mu_m$ and $d^2/\epsilon_0$ for electric dipoles of strength $d$, and $\theta$ is the angle between the dipole separation $\br$ and the polarization axis, which we take to be the $z$ direction. The particles also interact via a contact interaction of strength $g$ (note $g=0$ for spin-polarized fermions) and are taken to be confined within a trap  $\vt{\bx}$ of arbitrary geometry.

{\bf HLF theory:} 
 The single particle Wigner distribution function, within the semiclassical approximation,  is given by
\begin{align}
    W(\bx,\bk)=\frac{1}{e^{\beta[\epsilon(\bx,\bk)-\mu]}-\eta},\label{e:wigner}
\end{align} 
where $\eta=1$ for bosons and $\eta=-1$ for fermions,   $\mu$ is the chemical potential, and $\beta=1/\kt$ is the inverse temperature. 
The HLF theory is based on a trial dispersion relation 
\begin{align}
    \epsilon(\bx,\bk) &= \frac{\hbar^2}{2m}\left[\kappa_\rho(\bx) k_\rho^2 + \kappa_z(\bx) k_z^2\right] +\evar(\bx),\label{epsilon}
\end{align}
where $k_\rho\!=\!\sqrt{k_x^2+k_y^2}$, and the effective potential $\evar(\bx)$, as we show below,  includes the influence of trap, direct  and exchange interaction.  
We have also introduced local momentum distortion fields $\kappa_\rho(\bx)$ and $\kappa_z(\bx)$, which describe a spatially varying anisotropy of the momentum distribution with respect to the $z$ axis (i.e.~direction of dipole polarization). Because the momentum distortion determines the anisotropy of the pair correlation function \cite{Baillie2012a}, these fields parameterize the exchange interaction within the HLF theory. 
We note that the cylindrical symmetry of the DDI \eqref{e:udd} allows us to make the decomposition into $\kappa_\rho(\bx)$ and $\kappa_z(\bx)$ fields, irrespective of the trap geometry.  

Using the trial dispersion the position density is given by
\begin{align}
     n(\bx) &= \int \frac{d\bk}{(2\pi)^3} W(\bx,\bk) = \frac{\bosee[\eta]{3/2}{\beta[\mu-\evar(\bx)]}}{\ldb^3(\bx)}, \label{e:nx}
\end{align}
where $\ldb(\bx)\equiv \sqrt{2\pi \beta \hbar^2/m^*(\bx)}$ is the thermal de Broglie wavelength with spatially dependent effective mass $m^*(\bx)=m/\left[\kappa_\rho(\bx)^2\kappa_z(\bx)\right]^{1/3}$, and $\zeta^\eta_{\nu}(z)=\sum_{k=1}^{\infty}\eta^{k-1}z^k/k^\nu$ is the polylogarithm function.

By applying a variational principle to the free energy, we derive the equations for the local momentum distortion fields and the effective potential which define the HLF theory.
The exact equilibrium (grand) free energy $\Omega_{\mathrm{ex}}$ satisfies \cite{BlaizotRipka} 
\begin{equation}
\Omega_{\mathrm{ex}}\le  \Omega_{\mathrm{HLF}}\equiv \Omega_0-E_0+\langle \hat{H}\rangle_0,\label{Fvarprinc}
\end{equation}
where $\Omega_{\mathrm{HLF}}$ is the HLF free energy and 
\begin{align}
\Omega_0&\!=\! \int\! \frac{d\bx\,d\bk}{\beta(2\pi)^3}\eta\ln\left(1\!-\!\eta e^{\beta[\mu-\epsilon(\bx,\bk)]}\right)\!=\!-\!\int \!d\bx \, \tfrac{2}{3} K(\bx),\\
E_0&=\int \frac{d\bx\,d\bk}{ (2\pi)^3} \epsilon(\bx,\bk)W(\bx,\bk),\\
&=\int d\bx\left[  K(\bx) 
+  \evar(\bx)  n(\bx)\right],
\end{align}
are the free energy and single particle energy, respectively, with
\begin{equation}
 K(\bx)\equiv\frac{3}{2}\frac{\kt}{\ldb^3(\bx)}\bosee[\eta]{5/2}{\beta[\mu-\evar(\bx)]}.\label{Kx}
\end{equation}
The quantity  $\langle \hat{H}\rangle_0 = E_K\!+\!E_V\!+\! E_C\!+\!E_D\!+\!E_E$ is the HLF expectation of the Hamiltonian   \cite{Zhang2010a,Baillie2010b} with:
\begin{align}
     E_K &= \int \frac{d\bx\, d\bk}{(2\pi)^3} \frac{\hbar^2 k^2}{2m}W(\bx,\bk), \\
     E_V &= \int d\bx\, \vt{\bx}n(\bx),\\  
     E_C &= g\int d\bx\, n^2(\bx), \label{e:EC}\\
     E_D &= \half\int d\bx\, \Phi_D(\bx)n(\bx), \label{e:ECD}\\
     E_E &= \frac{\eta}{2}\int \frac{d\bx\, d\bk}{(2\pi)^3} \Phi_E (\bx,\bk) W(\bx,\bk).\label{e:EE}
\end{align}
 The contributions to $\langle{\hat H}\rangle_0$ are the kinetic energy $(E_K)$; the trap energy $(E_V)$; the combined direct and exchange contract interaction term ($E_C$); the direct dipolar term ($E_D$), with $\Phi_D(\bx)=\int d\bx'\,\UD(\bx-\bx') n(\bx')$;   the dipolar exchange interaction ($E_E$), where
 \begin{equation}
\Phi_E(\bx,\bk)= \int \frac{d\bk'}{(2\pi)^3}\,\UDt(\bk-\bk')W(\bx,\bk'),\label{phiefull}
\end{equation}
with $\UDt(\bk)=\Cdd (\cos^2\theta_\bk-\tfrac{1}{3})$  the Fourier transform of $\UD(\bx)$.
The expressions for the interaction terms, \eqref{e:EC} - \eqref{e:EE}, are obtained using HF factorization to decompose second order correlation functions into products of single particle correlation functions, which can be expressed in terms of the Wigner function. Evaluating the above expressions within the HLF ansatz yields
\begin{subequations}
\begin{align}
   \langle{\hat H}\rangle_0 & =  
    \int d\bx\!\left\{
   \left[
\frac{2}{3\kappa_\rho(\bx)} + \frac{1}{3\kappa_z(\bx)}
\right]K(\bx)\right.\label{H0kin}\\
     +&\left. \left[\vt{\bx}\!+ \!g n(\bx) \!+  \!\half \Phi_D(\bx) \!+ \!\frac{\eta}{2}\Phi_E(\bx)  \!\right] n(\bx)\!\right\}\!,\label{H0int}
\end{align}
  \end{subequations} 
with the local  exchange term  $\Phi_E(\bx)$ obtained from
\begin{align}
\hspace{-0.33cm}  \Phi_E(\bx) n(\bx)=\int \frac{ d\bk }{(2\pi)^3}\Phi_E(\bx,\bk)W(\bx,\bk).\label{e:EEx}
\end{align}
In addition to being local in position space,  $\Phi_E(\bx)$  has the simple analytic form
\begin{align}
    \Phi_E(\bx)\equiv  \Cdd  J\left[\delta(\bx)\right]  n(\bx),\label{PhiEn}
\end{align}
where $\delta(\bx)\equiv \kappa_z(\bx)/\kappa_\rho(\bx)-1$ is the relative distortion of the momentum distribution and\footnote{We note that $(\sinh^{-1}\sqrt{u})/\sqrt{u}$ is real for $u\ge-1$, that our $J(u)=I[(1+u)^{1/3}]/6$ of \cite{Sogo2009a}, and that $J(u)$ is easily differentiated for use in \eqref{e:deltafp} (see \cite{Baillie2012a}).}
\begin{align}
    J(u) &=  
    \left[\sqrt{1+u}
    \,(\sinh^{-1} \sqrt{u})/\sqrt{u} -1 \right]/u -\tfrac13.
\end{align}
is a monotonically decreasing function of $u$ with $J(0)=0$. Result (\ref{PhiEn}) shows that the effective exchange potential depends on the density and is only non-zero when the local momentum distribution is distorted from spherical symmetry [taking $\delta(\bx)=0$, $\Phi_E(\bx)$ and $E_E$ are zero and HLF reduces to Hartree theory]. The exchange potential appears with a pre-factor of $\eta$ in Eq.~\eqref{H0int} and we find that $\delta(\bx)>0$ for bosons and $\delta(\bx)<0$ for fermions so that $E_E$ is always negative.
The local form of exchange \eqref{PhiEn} we have arrived at is the central result that allows us to formulate a tractable and flexible theory. It is worth pausing to briefly compare to the HF treatment in which the full Wigner function needs to be evaluated and then convolved with the interaction potential  to obtain the exchange  potential (\ref{phiefull}) (e.g.~see \cite{Zhang2010a,Baillie2010b}). 
In contrast HLF theory does not require evaluating the  Wigner function, yet contains the momentum dependence of the exchange term  parameterized by our two position dependent distortion fields  $\{\kappa_\rho(\bx),\kappa_z(\bx)\}$ [or equivalently  $\{m^*(\bx),\delta(\bx)$\}].

{\bf HLF equations:} By requiring that $\Omega_{\mathrm{HLF}}$ (\ref{Fvarprinc})  is stationary with respect to arbitrary variations of $\evar(\bx)$, $\kappa_\rho(\bx)$ and $\kappa_z(\bx)$ we find: 
\begin{align}
 \! \!\!\evar(\bx)   &=  \vt{\bx} + 2g n(\bx) + \Phi_D(\bx) + \eta \Phi_E(\bx)\label{e:evaralt},\\
       \!\!\! \!\!\!\frac{m}{m^*(\bx)} &= \frac{1+\tfrac{2}{3}\delta(\bx)}{[1+\delta(\bx)]^{2/3}}\label{e:kappadelta},\\
    \delta(\bx) &= \!-\!\frac{9\eta\Cdd }{4}\frac{  n^2(\bx)}{K(\bx) } 
    [1\!+\!\tfrac{2}{3}\delta(\bx)][1\!+\!\delta(\bx)]J'[\delta(\bx)].\label{e:deltafp}
    \end{align}
Equation~(\ref{e:evaralt}) for the effective potential includes the local exchange potential.
The relative momentum distortion field $\delta(\bx)$ is determined by solving the transcendental Eq.~(\ref{e:deltafp}), and from this the  effective mass  is immediately given using Eq.~(\ref{e:kappadelta}). We note that  Eq.~(\ref{e:kappadelta}) ensures that the local kinetic energy is $K(\bx)$ [i.e.~the prefactor of $K(\bx)$ in Eq.~(\ref{H0kin}) is unity].

Equations~(\ref{e:evaralt})-(\ref{e:deltafp}), in conjunction with Eqs.~(\ref{e:nx}) and (\ref{Kx}), form the core set of equations of our theory that must be solved self-consistently. 
The direct potential, $\Phi_D(\bx)$, can be efficiently computed using the convolution theorem. We note that a number of accurate and efficient techniques for doing this have been developed for the purpose of solving the Gross-Pitaevskii equation with DDIs (e.g.~see \cite{Ronen2006a}). 

{\bf HGF equations:}
We can develop a simplified version of HLF by setting a single global distortion, implemented by ignoring the $\bx$ dependence of the momentum distortion fields. Minimizing the free energy we find Eqs.~\eqref{PhiEn}, \eqref{e:evaralt} and \eqref{e:kappadelta} (without position dependence of $\delta$ or $m^*$) and 
\begin{align}
    \delta = -\frac{9\eta\Cdd }{4}\frac{  \int d\bx \,n^2(\bx)}{\int d\bx \, K(\bx) } (1+\tfrac{2}{3}\delta)(1+\delta)J'(\delta),
\end{align}
which we refer to as the Hartree Global-Fock (HGF) theory. The HGF theory captures the average exchange effects, and thus provides a good description of quantities such as the position and momentum distributions.  For many predictions the HGF theory will be inaccurate because the relevant properties are determined by local properties, e.g.~mechanical stability is determined by  the densest part of the gas near trap center, where local exchange effects are largest and drive the collapse to occur at lower dipole strengths. Similar considerations will be important in predicting phase transitions. HLF is just as easy to implement as HGF and calculation times are similar, with HGF approximately twice as fast as HLF  calculations.

{\bf Results:} 
We validate the HLF theory by comparison to HF and Hartree calculations for a system in the harmonic trap
$
\vt\bx=\frac{1}{2}m\omega_{\rho}^2(x^2 + y^2 + \lambda^2z^2),
$ 
with   $\lambda=\omega_z/\omega_{\rho}$. To simplify our presentation we only discuss HGF calculations in cases that help illuminate its differences from HLF.

\begin{figure}[!tbh]
\begin{center}
\hspace*{-0.15in}\includegraphics[width=3.4in]{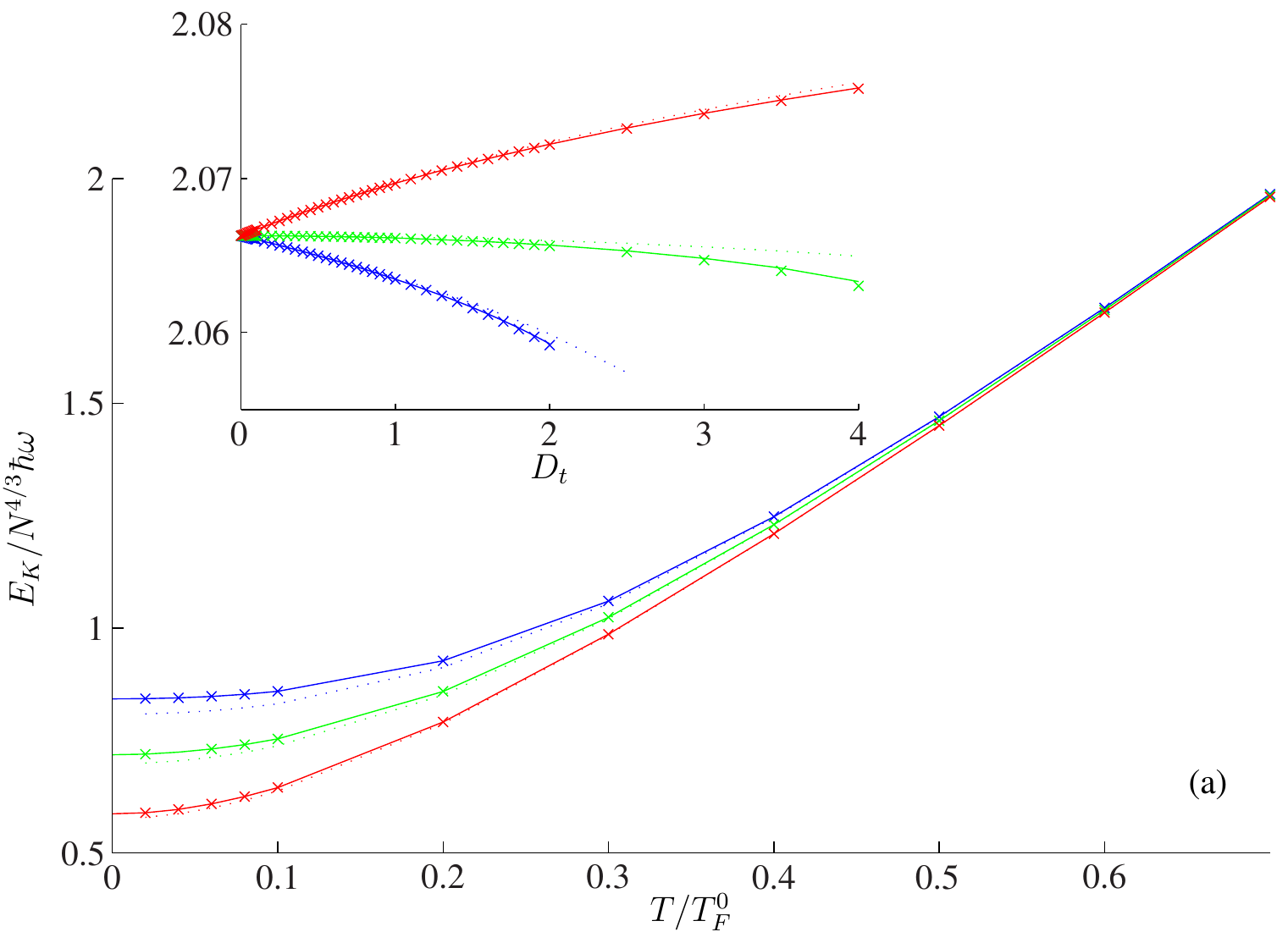}
\hspace*{-0.15in}\includegraphics[width=3.4in]{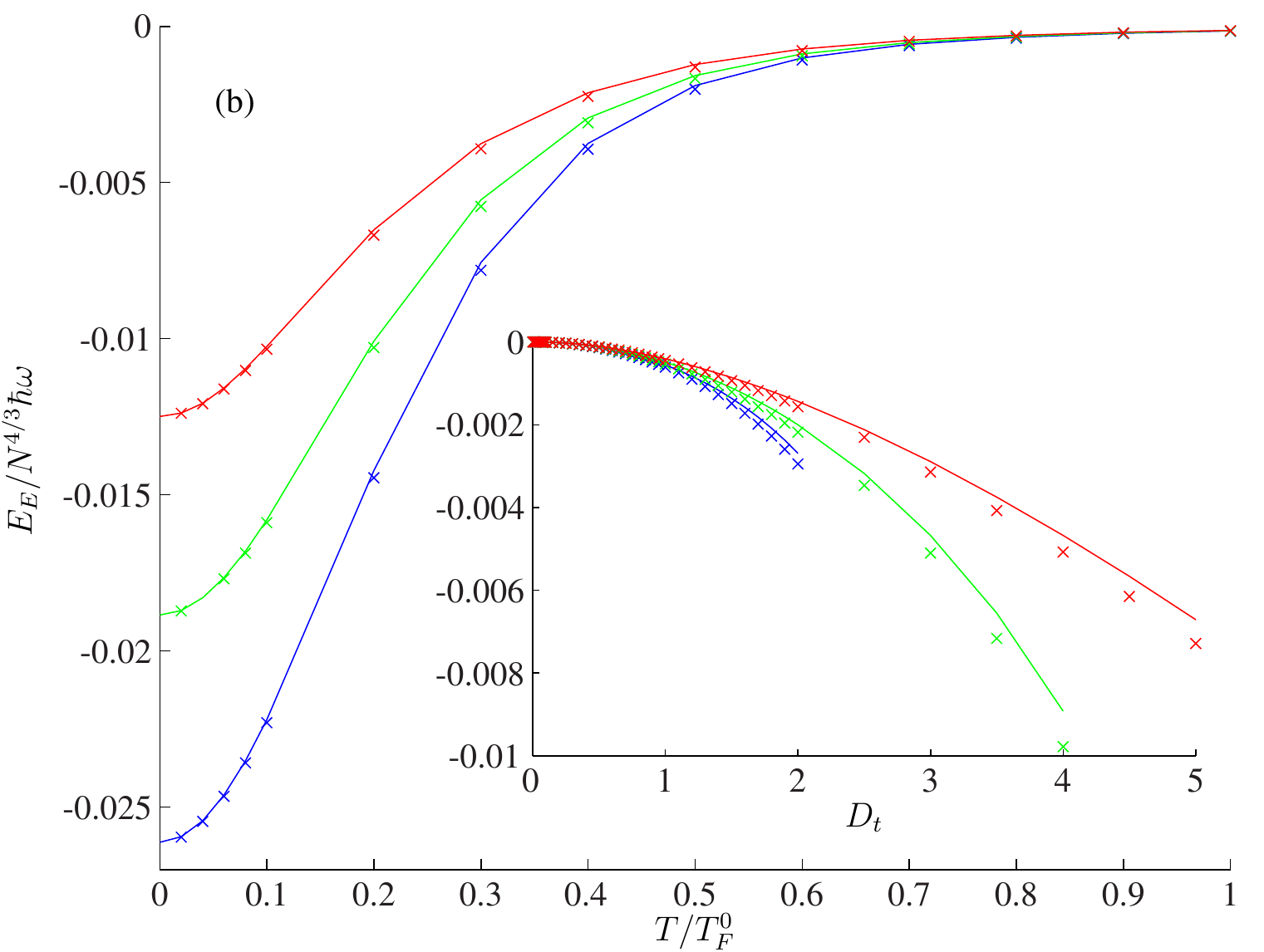}
\caption{(color online) Comparison of energies for pure ($g=0$) dipolar theories showing HF (crosses), HLF (solid curves), Hartree (dotted curves, zero for $E_E$ so not shown), for aspect ratio $\lambda=0.1$ (blue, dark grey), 1 (green, light grey), 10 (red, grey). (a) Kinetic energy $E_K$, (b) dipolar exchange energy $E_E$. Main figures are for fermions with $D_t=1$ and insets are for a Bose gas at $T=1.5T_c^0$.   
Dipole strength is parameterized in terms of  $D_t= \Cdd N^{1/6}/(4\pi\hbar\omega \aho^3)$, where $a_{\rm{ho}}=\sqrt{\hbar/m\omega}$  and $\omega=\sqrt[3]{\omega_{\rho}^2\omega_z}$, with $T_c^0 =\sqrt[3]{N/\zeta(3)}\hbar\omega/k_B$ and $T_F^0=\sqrt[3]{6N}\hbar\omega/k_B$ the ideal gas Bose-Einstein condensation and Fermi temperature, respectively. Results in the insets terminate at finite $D_t$ due to approaching instability \cite{Bisset2011}.\label{Fig:BFTherm}}
\end{center}\vspace{-0.4cm}
\end{figure}
For given $\mu$ and $T$ we find that the HLF free energy is above, but close to the full HF value, and appreciably lower than the Hartree value.  
In Fig.~\ref{Fig:BFTherm} we compare the kinetic and dipolar-exchange energy (both give  important contributions to $\Omega$) for Bose and Fermi systems with fixed mean number of particles $N$. For the kinetic energy we find that HF and HLF calculations are in excellent agreement, and discernibly different to the Hartree results. This difference, which is both positive and negative, arises directly from the momentum distortion as well as from the self-consistent effects of interactions changing the chemical potential.\footnote{At fixed $\mu$ and $T$, the Hartree value for $E_K$ is less than HF and HLF.}
 
The exchange energy $E_E$ is zero for the Hartree theory and  Fig.~\ref{Fig:BFTherm}(b) shows the predictions of HF and HLF theories, again revealing excellent agreement. In the $T\to0$ limit of the Fermi gas\footnote{For  $T\to0$ fermions in HLF  we use  
$ n\to\tfrac{\sqrt{2}}{3\pi^2}\left\{m^*[\mu-\evar]\right\}^{3/2}\!/\hbar^3$,  and
$K\to\tfrac{\sqrt{2}}{5\pi^2}(m^*)^{3/2}\left[\mu-\evar\right]^{5/2}\!/\hbar^3$. 
 This  limit is difficult to realize in HF calculations where the sharp Fermi surface in $W(\bx,\bk)$ (e.g.~see \cite{Zhang2009a}) is smeared by the numerical grid resolution revealing an additional advantage of HLF.
 } the direct and exchange contributions are of similar magnitude (i.e.~$|E_E|\approx |E_D|$) for spherically symmetric traps ($\lambda=1$),  and $|E_E|$ is about an order of magnitude smaller than $|E_D|$ for the anisotropic cases with $\lambda=10$ and $0.1$. This is because when the trap distorts the spatial distribution away from being nearly spherical the direct interaction (\ref{e:ECD}) is strongly enhanced, while $|E_E|$ remains roughly the same size.

\begin{figure}[!tbh]
\begin{center}
\hspace*{-0.15in}\includegraphics[width=3.6in]{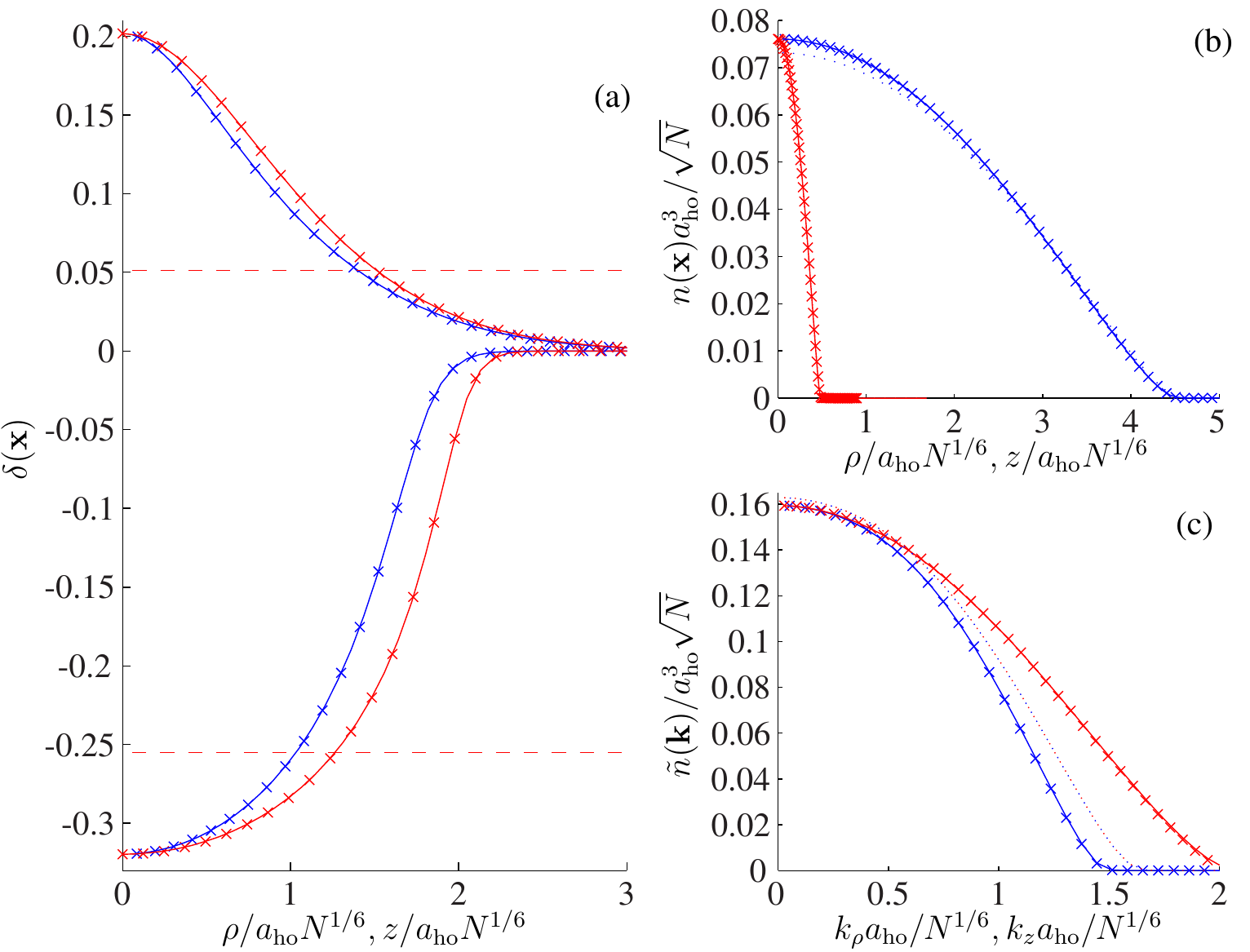} 
\caption{(color online) Comparison of pure ($g=0$) dipolar theories showing HF (crosses), HLF (solid curves), HGF (dashed lines) and Hartree (dotted curves) in the radial (blue, dark grey) and axial (red, grey) directions. 
(a) $\delta(\bx)$ (Hartree results are not shown, being all zero) with $\lambda=1$ and $D_t = 1$ for bosons at $T=1.2T_c^0$ (upper curves) and fermions at $T=0.1T_F^0$ (lower curves). (b) $n(\bx)$ and (c) momentum density $\tilde{n}(\bk) = \int \frac{d\bx}{(2\pi)^3} W(\bx,\bk)$ for fermions at $T=0.02T_F^0$ with $\lambda=10$ and $D_t=2$. The HGF agreement in (b) and (c) is reasonable and is not shown.
 \label{Fig:localmtmdistort}}
\end{center}\vspace{-0.4cm}
\end{figure}

The field $\delta(\bx)$, which is a key element of the HLF theory,  is a measure of a local quadrupolar moment (i.e.~distortion)  of the momentum distribution, given by
\begin{equation}
\delta(\bx) \equiv  {\frac{\gamma_{k_x}(\bx)}{\gamma_{k_z}(\bx)}}-1,\quad \gamma_{\nu}(\bx)\equiv\int \frac{d\bk}{(2\pi)^3}\,\nu^2W(\bx,\bk),\label{distort}\end{equation}
where $\{\gamma_{k_x}(\bx),\gamma_{k_z}(\bx)\}$ are the local momentum moments.
We can use (\ref{distort}) to evaluate $\delta(\bx)$ from the full HF solutions [recall $\delta(\bx)=0$ in the Hartree theory]. In Fig.~\ref{Fig:localmtmdistort}(a) we show the local momentum distortion for Bose and Fermi systems. 
Our results demonstrate that the momentum distortion varies spatially, with the largest distortion occurring at trap center (i.e.~where density is highest), and that this effect is accurately captured by HLF theory. 
We also show HGF results which demonstrate that this approach predicts a reasonable average distortion, but fails to capture its spatial dependence.

Both the Bose and Fermi systems exhibit similar behavior in their position-space distortion effects, i.e.~the density elongates along the polarization ($z$) direction to reduce  $E_D$. The momentum space distortion \cite{Baillie2012a} is distinctive: to reduce $E_E$ the Fermi system elongates along the $k_z$ direction whereas the Bose system reduces its $k_z$ extent to instead expand in the radial momentum plane. 
This behavior is also apparent in the short range correlations between particles (e.g.~see discussion in \cite{Baillie2012a}) and should be verifiable in current experiments \cite{Donner2007a}. 
Density profiles in position and momentum space in Fig.~\ref{Fig:localmtmdistort}(b) and (c), respectively, show that while the Hartree position density is in reasonable agreement with HF and HLF, the Hartree theory fails to capture the difference between the momentum density in radial and axial directions.

{\bf Conclusion and outlook:} 
In this paper we have introduced a variational ansatz that converts the HF theory of dipolar Bose and Fermi gases to a local density dependent theory, with negligible error compared to the full HF solutions.   
 The resulting calculations are practical to undertake with a dramatic reduction in required computing resources compared to HF and should support this burgeoning field of dipolar quantum gases.  This approach also provides insight into the manifestation of exchange effects in the dipolar gas, such as local and global momentum distortion, that could be verified in current experiments. In future work we will extend this approach to superfluid Bose and Fermi gases.

{\bf Acknowledgments:}  This work was supported by the Marsden Fund of New Zealand  contract UOO0924.


\begin{thebibliography}{52}%
\makeatletter
\providecommand \@ifxundefined [1]{%
 \@ifx{#1\undefined}
}%
\providecommand \@ifnum [1]{%
 \ifnum #1\expandafter \@firstoftwo
 \else \expandafter \@secondoftwo
 \fi
}%
\providecommand \@ifx [1]{%
 \ifx #1\expandafter \@firstoftwo
 \else \expandafter \@secondoftwo
 \fi
}%
\providecommand \natexlab [1]{#1}%
\providecommand \enquote  [1]{``#1''}%
\providecommand \bibnamefont  [1]{#1}%
\providecommand \bibfnamefont [1]{#1}%
\providecommand \citenamefont [1]{#1}%
\providecommand \href@noop [0]{\@secondoftwo}%
\providecommand \href [0]{\begingroup \@sanitize@url \@href}%
\providecommand \@href[1]{\@@startlink{#1}\@@href}%
\providecommand \@@href[1]{\endgroup#1\@@endlink}%
\providecommand \@sanitize@url [0]{\catcode `\\12\catcode `\$12\catcode
  `\&12\catcode `\#12\catcode `\^12\catcode `\_12\catcode `\%12\relax}%
\providecommand \@@startlink[1]{}%
\providecommand \@@endlink[0]{}%
\providecommand \url  [0]{\begingroup\@sanitize@url \@url }%
\providecommand \@url [1]{\endgroup\@href {#1}{\urlprefix }}%
\providecommand \urlprefix  [0]{URL }%
\providecommand \Eprint [0]{\href }%
\providecommand \doibase [0]{http://dx.doi.org/}%
\providecommand \selectlanguage [0]{\@gobble}%
\providecommand \bibinfo  [0]{\@secondoftwo}%
\providecommand \bibfield  [0]{\@secondoftwo}%
\providecommand \translation [1]{[#1]}%
\providecommand \BibitemOpen [0]{}%
\providecommand \bibitemStop [0]{}%
\providecommand \bibitemNoStop [0]{.\EOS\space}%
\providecommand \EOS [0]{\spacefactor3000\relax}%
\providecommand \BibitemShut  [1]{\csname bibitem#1\endcsname}%
\let\auto@bib@innerbib\@empty
\bibitem [{\citenamefont {Griesmaier}\ \emph {et~al.}(2005)\citenamefont
  {Griesmaier}, \citenamefont {Werner}, \citenamefont {Hensler}, \citenamefont
  {Stuhler},\ and\ \citenamefont {Pfau}}]{Griesmaier2005a}%
  \BibitemOpen
  \bibfield  {author} {\bibinfo {author} {\bibfnamefont {A.}~\bibnamefont
  {Griesmaier}}, \bibinfo {author} {\bibfnamefont {J.}~\bibnamefont {Werner}},
  \bibinfo {author} {\bibfnamefont {S.}~\bibnamefont {Hensler}}, \bibinfo
  {author} {\bibfnamefont {J.}~\bibnamefont {Stuhler}}, \ and\ \bibinfo
  {author} {\bibfnamefont {T.}~\bibnamefont {Pfau}},\ }\href {\doibase
  10.1103/PhysRevLett.94.160401} {\bibfield  {journal} {\bibinfo  {journal}
  {Phys. Rev. Lett.}\ }\textbf {\bibinfo {volume} {94}},\ \bibinfo {pages}
  {160401} (\bibinfo {year} {2005})}\BibitemShut {NoStop}%
\bibitem [{\citenamefont {Bismut}\ \emph {et~al.}(2010)\citenamefont {Bismut},
  \citenamefont {Pasquiou}, \citenamefont {Mar\'echal}, \citenamefont {Pedri},
  \citenamefont {Vernac}, \citenamefont {Gorceix},\ and\ \citenamefont
  {Laburthe-Tolra}}]{Bismut2010a}%
  \BibitemOpen
  \bibfield  {author} {\bibinfo {author} {\bibfnamefont {G.}~\bibnamefont
  {Bismut}}, \bibinfo {author} {\bibfnamefont {B.}~\bibnamefont {Pasquiou}},
  \bibinfo {author} {\bibfnamefont {E.}~\bibnamefont {Mar\'echal}}, \bibinfo
  {author} {\bibfnamefont {P.}~\bibnamefont {Pedri}}, \bibinfo {author}
  {\bibfnamefont {L.}~\bibnamefont {Vernac}}, \bibinfo {author} {\bibfnamefont
  {O.}~\bibnamefont {Gorceix}}, \ and\ \bibinfo {author} {\bibfnamefont
  {B.}~\bibnamefont {Laburthe-Tolra}},\ }\href {\doibase
  10.1103/PhysRevLett.105.040404} {\bibfield  {journal} {\bibinfo  {journal}
  {Phys. Rev. Lett.}\ }\textbf {\bibinfo {volume} {105}},\ \bibinfo {pages}
  {040404} (\bibinfo {year} {2010})}\BibitemShut {NoStop}%
\bibitem [{\citenamefont {Pasquiou}\ \emph {et~al.}(2011)\citenamefont
  {Pasquiou}, \citenamefont {Bismut}, \citenamefont {Mar\'echal}, \citenamefont
  {Pedri}, \citenamefont {Vernac}, \citenamefont {Gorceix},\ and\ \citenamefont
  {Laburthe-Tolra}}]{Pasquiou2011a}%
  \BibitemOpen
  \bibfield  {author} {\bibinfo {author} {\bibfnamefont {B.}~\bibnamefont
  {Pasquiou}}, \bibinfo {author} {\bibfnamefont {G.}~\bibnamefont {Bismut}},
  \bibinfo {author} {\bibfnamefont {E.}~\bibnamefont {Mar\'echal}}, \bibinfo
  {author} {\bibfnamefont {P.}~\bibnamefont {Pedri}}, \bibinfo {author}
  {\bibfnamefont {L.}~\bibnamefont {Vernac}}, \bibinfo {author} {\bibfnamefont
  {O.}~\bibnamefont {Gorceix}}, \ and\ \bibinfo {author} {\bibfnamefont
  {B.}~\bibnamefont {Laburthe-Tolra}},\ }\href {\doibase
  10.1103/PhysRevLett.106.015301} {\bibfield  {journal} {\bibinfo  {journal}
  {Phys. Rev. Lett.}\ }\textbf {\bibinfo {volume} {106}},\ \bibinfo {pages}
  {015301} (\bibinfo {year} {2011})}\BibitemShut {NoStop}%
\bibitem [{\citenamefont {Lu}\ \emph {et~al.}(2011)\citenamefont {Lu},
  \citenamefont {Burdick}, \citenamefont {Youn},\ and\ \citenamefont
  {Lev}}]{Mingwu2011a}%
  \BibitemOpen
  \bibfield  {author} {\bibinfo {author} {\bibfnamefont {M.}~\bibnamefont
  {Lu}}, \bibinfo {author} {\bibfnamefont {N.~Q.}\ \bibnamefont {Burdick}},
  \bibinfo {author} {\bibfnamefont {S.~H.}\ \bibnamefont {Youn}}, \ and\
  \bibinfo {author} {\bibfnamefont {B.~L.}\ \bibnamefont {Lev}},\ }\href
  {\doibase 10.1103/PhysRevLett.107.190401} {\bibfield  {journal} {\bibinfo
  {journal} {Phys. Rev. Lett.}\ }\textbf {\bibinfo {volume} {107}},\ \bibinfo
  {pages} {190401} (\bibinfo {year} {2011})}\BibitemShut {NoStop}%
\bibitem [{\citenamefont {Aikawa}\ \emph {et~al.}(2012)\citenamefont {Aikawa},
  \citenamefont {Frisch}, \citenamefont {Mark}, \citenamefont {Baier},
  \citenamefont {Rietzler}, \citenamefont {Grimm},\ and\ \citenamefont
  {Ferlaino}}]{Aikawa2012a}%
  \BibitemOpen
  \bibfield  {author} {\bibinfo {author} {\bibfnamefont {K.}~\bibnamefont
  {Aikawa}}, \bibinfo {author} {\bibfnamefont {A.}~\bibnamefont {Frisch}},
  \bibinfo {author} {\bibfnamefont {M.}~\bibnamefont {Mark}}, \bibinfo {author}
  {\bibfnamefont {S.}~\bibnamefont {Baier}}, \bibinfo {author} {\bibfnamefont
  {A.}~\bibnamefont {Rietzler}}, \bibinfo {author} {\bibfnamefont
  {R.}~\bibnamefont {Grimm}}, \ and\ \bibinfo {author} {\bibfnamefont
  {F.}~\bibnamefont {Ferlaino}},\ }\href {\doibase
  10.1103/PhysRevLett.108.210401} {\bibfield  {journal} {\bibinfo  {journal}
  {Phys. Rev. Lett.}\ }\textbf {\bibinfo {volume} {108}},\ \bibinfo {pages}
  {210401} (\bibinfo {year} {2012})}\BibitemShut {NoStop}%
\bibitem [{\citenamefont {Lu}\ \emph {et~al.}(2012)\citenamefont {Lu},
  \citenamefont {Burdick},\ and\ \citenamefont {Lev}}]{Lu2012a}%
  \BibitemOpen
  \bibfield  {author} {\bibinfo {author} {\bibfnamefont {M.}~\bibnamefont
  {Lu}}, \bibinfo {author} {\bibfnamefont {N.~Q.}\ \bibnamefont {Burdick}}, \
  and\ \bibinfo {author} {\bibfnamefont {B.~L.}\ \bibnamefont {Lev}},\ }\href
  {\doibase 10.1103/PhysRevLett.108.215301} {\bibfield  {journal} {\bibinfo
  {journal} {Phys. Rev. Lett.}\ }\textbf {\bibinfo {volume} {108}},\ \bibinfo
  {pages} {215301} (\bibinfo {year} {2012})}\BibitemShut {NoStop}%
\bibitem [{\citenamefont {Aikawa}\ \emph {et~al.}(2010)\citenamefont {Aikawa},
  \citenamefont {Akamatsu}, \citenamefont {Hayashi}, \citenamefont {Oasa},
  \citenamefont {Kobayashi}, \citenamefont {Naidon}, \citenamefont {Kishimoto},
  \citenamefont {Ueda},\ and\ \citenamefont {Inouye}}]{Aikawa2010a}%
  \BibitemOpen
  \bibfield  {author} {\bibinfo {author} {\bibfnamefont {K.}~\bibnamefont
  {Aikawa}}, \bibinfo {author} {\bibfnamefont {D.}~\bibnamefont {Akamatsu}},
  \bibinfo {author} {\bibfnamefont {M.}~\bibnamefont {Hayashi}}, \bibinfo
  {author} {\bibfnamefont {K.}~\bibnamefont {Oasa}}, \bibinfo {author}
  {\bibfnamefont {J.}~\bibnamefont {Kobayashi}}, \bibinfo {author}
  {\bibfnamefont {P.}~\bibnamefont {Naidon}}, \bibinfo {author} {\bibfnamefont
  {T.}~\bibnamefont {Kishimoto}}, \bibinfo {author} {\bibfnamefont
  {M.}~\bibnamefont {Ueda}}, \ and\ \bibinfo {author} {\bibfnamefont
  {S.}~\bibnamefont {Inouye}},\ }\href {\doibase
  10.1103/PhysRevLett.105.203001} {\bibfield  {journal} {\bibinfo  {journal}
  {Phys. Rev. Lett.}\ }\textbf {\bibinfo {volume} {105}},\ \bibinfo {pages}
  {203001} (\bibinfo {year} {2010})}\BibitemShut {NoStop}%
\bibitem [{\citenamefont {Ni}\ \emph {et~al.}(2008)\citenamefont {Ni},
  \citenamefont {Ospelkaus}, \citenamefont {de~Miranda}, \citenamefont {Pe'er},
  \citenamefont {Neyenhuis}, \citenamefont {Zirbel}, \citenamefont
  {Kotochigova}, \citenamefont {Julienne}, \citenamefont {Jin},\ and\
  \citenamefont {Ye}}]{Ni2008a}%
  \BibitemOpen
  \bibfield  {author} {\bibinfo {author} {\bibfnamefont {K.-K.}\ \bibnamefont
  {Ni}}, \bibinfo {author} {\bibfnamefont {S.}~\bibnamefont {Ospelkaus}},
  \bibinfo {author} {\bibfnamefont {M.~H.~G.}\ \bibnamefont {de~Miranda}},
  \bibinfo {author} {\bibfnamefont {A.}~\bibnamefont {Pe'er}}, \bibinfo
  {author} {\bibfnamefont {B.}~\bibnamefont {Neyenhuis}}, \bibinfo {author}
  {\bibfnamefont {J.~J.}\ \bibnamefont {Zirbel}}, \bibinfo {author}
  {\bibfnamefont {S.}~\bibnamefont {Kotochigova}}, \bibinfo {author}
  {\bibfnamefont {P.~S.}\ \bibnamefont {Julienne}}, \bibinfo {author}
  {\bibfnamefont {D.~S.}\ \bibnamefont {Jin}}, \ and\ \bibinfo {author}
  {\bibfnamefont {J.}~\bibnamefont {Ye}},\ }\href@noop {} {\bibfield  {journal}
  {\bibinfo  {journal} {Science}\ }\textbf {\bibinfo {volume} {322}},\ \bibinfo
  {pages} {231} (\bibinfo {year} {2008})}\BibitemShut {NoStop}%
\bibitem [{\citenamefont {Lahaye}\ \emph {et~al.}(2007)\citenamefont {Lahaye},
  \citenamefont {Koch}, \citenamefont {Fr{\"o}hlich}, \citenamefont {Fattori},
  \citenamefont {Metz}, \citenamefont {Griesmaier}, \citenamefont
  {Giovanazzi},\ and\ \citenamefont {Pfau}}]{Lahaye2007a}%
  \BibitemOpen
  \bibfield  {author} {\bibinfo {author} {\bibfnamefont {T.}~\bibnamefont
  {Lahaye}}, \bibinfo {author} {\bibfnamefont {T.}~\bibnamefont {Koch}},
  \bibinfo {author} {\bibfnamefont {B.}~\bibnamefont {Fr{\"o}hlich}}, \bibinfo
  {author} {\bibfnamefont {M.}~\bibnamefont {Fattori}}, \bibinfo {author}
  {\bibfnamefont {J.}~\bibnamefont {Metz}}, \bibinfo {author} {\bibfnamefont
  {A.}~\bibnamefont {Griesmaier}}, \bibinfo {author} {\bibfnamefont
  {S.}~\bibnamefont {Giovanazzi}}, \ and\ \bibinfo {author} {\bibfnamefont
  {T.}~\bibnamefont {Pfau}},\ }\href@noop {} {\bibfield  {journal} {\bibinfo
  {journal} {Nature}\ }\textbf {\bibinfo {volume} {448}},\ \bibinfo {pages}
  {672} (\bibinfo {year} {2007})}\BibitemShut {NoStop}%
\bibitem [{\citenamefont {G\'oral}\ \emph {et~al.}(2000)\citenamefont
  {G\'oral}, \citenamefont {Rza\ifmmode \mbox{\c{}}\else
  \c{}\fi{}\ifmmode~\dot{z}\else \.{z}\fi{}ewski},\ and\ \citenamefont
  {Pfau}}]{Goral2000a}%
  \BibitemOpen
  \bibfield  {author} {\bibinfo {author} {\bibfnamefont {K.}~\bibnamefont
  {G\'oral}}, \bibinfo {author} {\bibfnamefont {K.}~\bibnamefont {Rza\ifmmode
  \mbox{\c{}}\else \c{}\fi{}\ifmmode~\dot{z}\else \.{z}\fi{}ewski}}, \ and\
  \bibinfo {author} {\bibfnamefont {T.}~\bibnamefont {Pfau}},\ }\href {\doibase
  10.1103/PhysRevA.61.051601} {\bibfield  {journal} {\bibinfo  {journal} {Phys.
  Rev. A}\ }\textbf {\bibinfo {volume} {61}},\ \bibinfo {pages} {051601(R)}
  (\bibinfo {year} {2000})}\BibitemShut {NoStop}%
\bibitem [{\citenamefont {Kawaguchi}\ \emph
  {et~al.}(2006{\natexlab{a}})\citenamefont {Kawaguchi}, \citenamefont
  {Saito},\ and\ \citenamefont {Ueda}}]{Kawaguchi2006a}%
  \BibitemOpen
  \bibfield  {author} {\bibinfo {author} {\bibfnamefont {Y.}~\bibnamefont
  {Kawaguchi}}, \bibinfo {author} {\bibfnamefont {H.}~\bibnamefont {Saito}}, \
  and\ \bibinfo {author} {\bibfnamefont {M.}~\bibnamefont {Ueda}},\ }\href
  {\doibase 10.1103/PhysRevLett.96.080405} {\bibfield  {journal} {\bibinfo
  {journal} {Phys. Rev. Lett.}\ }\textbf {\bibinfo {volume} {96}},\ \bibinfo
  {eid} {080405} (\bibinfo {year} {2006}{\natexlab{a}})}\BibitemShut {NoStop}%
\bibitem [{\citenamefont {Ronen}\ \emph
  {et~al.}(2006{\natexlab{a}})\citenamefont {Ronen}, \citenamefont
  {Bortolotti}, \citenamefont {Blume},\ and\ \citenamefont
  {Bohn}}]{Ronen2006b}%
  \BibitemOpen
  \bibfield  {author} {\bibinfo {author} {\bibfnamefont {S.}~\bibnamefont
  {Ronen}}, \bibinfo {author} {\bibfnamefont {D.~C.~E.}\ \bibnamefont
  {Bortolotti}}, \bibinfo {author} {\bibfnamefont {D.}~\bibnamefont {Blume}}, \
  and\ \bibinfo {author} {\bibfnamefont {J.~L.}\ \bibnamefont {Bohn}},\ }\href
  {\doibase 10.1103/PhysRevA.74.033611} {\bibfield  {journal} {\bibinfo
  {journal} {Phys. Rev. A}\ }\textbf {\bibinfo {volume} {74}},\ \bibinfo {eid}
  {033611} (\bibinfo {year} {2006}{\natexlab{a}})}\BibitemShut {NoStop}%
\bibitem [{\citenamefont {Kawaguchi}\ \emph
  {et~al.}(2006{\natexlab{b}})\citenamefont {Kawaguchi}, \citenamefont
  {Saito},\ and\ \citenamefont {Ueda}}]{Kawaguchi2006b}%
  \BibitemOpen
  \bibfield  {author} {\bibinfo {author} {\bibfnamefont {Y.}~\bibnamefont
  {Kawaguchi}}, \bibinfo {author} {\bibfnamefont {H.}~\bibnamefont {Saito}}, \
  and\ \bibinfo {author} {\bibfnamefont {M.}~\bibnamefont {Ueda}},\ }\href
  {\doibase 10.1103/PhysRevLett.97.130404} {\bibfield  {journal} {\bibinfo
  {journal} {Phys. Rev. Lett.}\ }\textbf {\bibinfo {volume} {97}},\ \bibinfo
  {eid} {130404} (\bibinfo {year} {2006}{\natexlab{b}})}\BibitemShut {NoStop}%
\bibitem [{\citenamefont {Yi}\ and\ \citenamefont {Pu}(2006)}]{Yi2006a}%
  \BibitemOpen
  \bibfield  {author} {\bibinfo {author} {\bibfnamefont {S.}~\bibnamefont
  {Yi}}\ and\ \bibinfo {author} {\bibfnamefont {H.}~\bibnamefont {Pu}},\ }\href
  {\doibase 10.1103/PhysRevA.73.061602} {\bibfield  {journal} {\bibinfo
  {journal} {Phys. Rev. A}\ }\textbf {\bibinfo {volume} {73}},\ \bibinfo
  {pages} {061602(R)} (\bibinfo {year} {2006})}\BibitemShut {NoStop}%
\bibitem [{\citenamefont {Kawaguchi}\ \emph {et~al.}(2007)\citenamefont
  {Kawaguchi}, \citenamefont {Saito},\ and\ \citenamefont
  {Ueda}}]{Kawaguchi2007a}%
  \BibitemOpen
  \bibfield  {author} {\bibinfo {author} {\bibfnamefont {Y.}~\bibnamefont
  {Kawaguchi}}, \bibinfo {author} {\bibfnamefont {H.}~\bibnamefont {Saito}}, \
  and\ \bibinfo {author} {\bibfnamefont {M.}~\bibnamefont {Ueda}},\ }\href
  {\doibase 10.1103/PhysRevLett.98.110406} {\bibfield  {journal} {\bibinfo
  {journal} {Phys. Rev. Lett.}\ }\textbf {\bibinfo {volume} {98}},\ \bibinfo
  {eid} {110406} (\bibinfo {year} {2007})}\BibitemShut {NoStop}%
\bibitem [{\citenamefont {Wilson}\ \emph {et~al.}(2009)\citenamefont {Wilson},
  \citenamefont {Ronen},\ and\ \citenamefont {Bohn}}]{Wilson2009a}%
  \BibitemOpen
  \bibfield  {author} {\bibinfo {author} {\bibfnamefont {R.~M.}\ \bibnamefont
  {Wilson}}, \bibinfo {author} {\bibfnamefont {S.}~\bibnamefont {Ronen}}, \
  and\ \bibinfo {author} {\bibfnamefont {J.~L.}\ \bibnamefont {Bohn}},\ }\href
  {\doibase 10.1103/PhysRevA.79.013621} {\bibfield  {journal} {\bibinfo
  {journal} {Phys. Rev. A}\ }\textbf {\bibinfo {volume} {79}},\ \bibinfo {eid}
  {013621} (\bibinfo {year} {2009})}\BibitemShut {NoStop}%
\bibitem [{\citenamefont {Parker}\ \emph {et~al.}(2009)\citenamefont {Parker},
  \citenamefont {Ticknor}, \citenamefont {Martin},\ and\ \citenamefont
  {O'Dell}}]{Parker2009a}%
  \BibitemOpen
  \bibfield  {author} {\bibinfo {author} {\bibfnamefont {N.~G.}\ \bibnamefont
  {Parker}}, \bibinfo {author} {\bibfnamefont {C.}~\bibnamefont {Ticknor}},
  \bibinfo {author} {\bibfnamefont {A.~M.}\ \bibnamefont {Martin}}, \ and\
  \bibinfo {author} {\bibfnamefont {D.~H.~J.}\ \bibnamefont {O'Dell}},\ }\href
  {\doibase 10.1103/PhysRevA.79.013617} {\bibfield  {journal} {\bibinfo
  {journal} {Phys. Rev. A}\ }\textbf {\bibinfo {volume} {79}},\ \bibinfo {eid}
  {013617} (\bibinfo {year} {2009})}\BibitemShut {NoStop}%
\bibitem [{\citenamefont {O'Dell}\ \emph {et~al.}(2004)\citenamefont {O'Dell},
  \citenamefont {Giovanazzi},\ and\ \citenamefont {Eberlein}}]{ODell2004a}%
  \BibitemOpen
  \bibfield  {author} {\bibinfo {author} {\bibfnamefont {D.~H.~J.}\
  \bibnamefont {O'Dell}}, \bibinfo {author} {\bibfnamefont {S.}~\bibnamefont
  {Giovanazzi}}, \ and\ \bibinfo {author} {\bibfnamefont {C.}~\bibnamefont
  {Eberlein}},\ }\href {\doibase 10.1103/PhysRevLett.92.250401} {\bibfield
  {journal} {\bibinfo  {journal} {Phys. Rev. Lett.}\ }\textbf {\bibinfo
  {volume} {92}},\ \bibinfo {pages} {250401} (\bibinfo {year}
  {2004})}\BibitemShut {NoStop}%
\bibitem [{\citenamefont {Zhang}\ and\ \citenamefont {Yi}(2009)}]{Zhang2009a}%
  \BibitemOpen
  \bibfield  {author} {\bibinfo {author} {\bibfnamefont {J.-N.}\ \bibnamefont
  {Zhang}}\ and\ \bibinfo {author} {\bibfnamefont {S.}~\bibnamefont {Yi}},\
  }\href {\doibase 10.1103/PhysRevA.80.053614} {\bibfield  {journal} {\bibinfo
  {journal} {Phys. Rev. A}\ }\textbf {\bibinfo {volume} {80}},\ \bibinfo
  {pages} {053614} (\bibinfo {year} {2009})}\BibitemShut {NoStop}%
\bibitem [{\citenamefont {Zhang}\ and\ \citenamefont {Yi}(2010)}]{Zhang2010a}%
  \BibitemOpen
  \bibfield  {author} {\bibinfo {author} {\bibfnamefont {J.-N.}\ \bibnamefont
  {Zhang}}\ and\ \bibinfo {author} {\bibfnamefont {S.}~\bibnamefont {Yi}},\
  }\href {\doibase 10.1103/PhysRevA.81.033617} {\bibfield  {journal} {\bibinfo
  {journal} {Phys. Rev. A}\ }\textbf {\bibinfo {volume} {81}},\ \bibinfo
  {pages} {033617} (\bibinfo {year} {2010})}\BibitemShut {NoStop}%
\bibitem [{\citenamefont {Baillie}\ and\ \citenamefont
  {Blakie}(2010)}]{Baillie2010b}%
  \BibitemOpen
  \bibfield  {author} {\bibinfo {author} {\bibfnamefont {D.}~\bibnamefont
  {Baillie}}\ and\ \bibinfo {author} {\bibfnamefont {P.~B.}\ \bibnamefont
  {Blakie}},\ }\href {\doibase 10.1103/PhysRevA.82.033605} {\bibfield
  {journal} {\bibinfo  {journal} {Phys. Rev. A}\ }\textbf {\bibinfo {volume}
  {82}},\ \bibinfo {pages} {033605} (\bibinfo {year} {2010})}\BibitemShut
  {NoStop}%
\bibitem [{\citenamefont {{Baillie}}\ and\ \citenamefont
  {{Blakie}}(2012)}]{Baillie2012a}%
  \BibitemOpen
  \bibfield  {author} {\bibinfo {author} {\bibfnamefont {D.}~\bibnamefont
  {{Baillie}}}\ and\ \bibinfo {author} {\bibfnamefont {P.~B.}\ \bibnamefont
  {{Blakie}}},\ }\href@noop {} {\bibfield  {journal} {\bibinfo  {journal}
  {ArXiv e-prints}\ } (\bibinfo {year} {2012})},\ \Eprint
  {http://arxiv.org/abs/1207.3153} {arXiv:1207.3153 [to appear in
  Phys.~Rev.~A]} \BibitemShut {NoStop}%
\bibitem [{\citenamefont {Ticknor}(2012)}]{Ticknor2012a}%
  \BibitemOpen
  \bibfield  {author} {\bibinfo {author} {\bibfnamefont {C.}~\bibnamefont
  {Ticknor}},\ }\href {\doibase 10.1103/PhysRevA.85.033629} {\bibfield
  {journal} {\bibinfo  {journal} {Phys. Rev. A}\ }\textbf {\bibinfo {volume}
  {85}},\ \bibinfo {pages} {033629} (\bibinfo {year} {2012})}\BibitemShut
  {NoStop}%
\bibitem [{\citenamefont {Miyakawa}\ \emph {et~al.}(2008)\citenamefont
  {Miyakawa}, \citenamefont {Sogo},\ and\ \citenamefont {Pu}}]{Miyakawa2008a}%
  \BibitemOpen
  \bibfield  {author} {\bibinfo {author} {\bibfnamefont {T.}~\bibnamefont
  {Miyakawa}}, \bibinfo {author} {\bibfnamefont {T.}~\bibnamefont {Sogo}}, \
  and\ \bibinfo {author} {\bibfnamefont {H.}~\bibnamefont {Pu}},\ }\href
  {\doibase 10.1103/PhysRevA.77.061603} {\bibfield  {journal} {\bibinfo
  {journal} {Phys. Rev. A}\ }\textbf {\bibinfo {volume} {77}},\ \bibinfo
  {pages} {061603(R)} (\bibinfo {year} {2008})}\BibitemShut {NoStop}%
\bibitem [{\citenamefont {Endo}\ \emph {et~al.}(2010)\citenamefont {Endo},
  \citenamefont {Miyakawa},\ and\ \citenamefont {Nikuni}}]{Endo2010a}%
  \BibitemOpen
  \bibfield  {author} {\bibinfo {author} {\bibfnamefont {Y.}~\bibnamefont
  {Endo}}, \bibinfo {author} {\bibfnamefont {T.}~\bibnamefont {Miyakawa}}, \
  and\ \bibinfo {author} {\bibfnamefont {T.}~\bibnamefont {Nikuni}},\ }\href
  {\doibase 10.1103/PhysRevA.81.063624} {\bibfield  {journal} {\bibinfo
  {journal} {Phys. Rev. A}\ }\textbf {\bibinfo {volume} {81}},\ \bibinfo
  {pages} {063624} (\bibinfo {year} {2010})}\BibitemShut {NoStop}%
\bibitem [{\citenamefont {Lima}\ and\ \citenamefont
  {Pelster}(2010)}]{Lima2010a}%
  \BibitemOpen
  \bibfield  {author} {\bibinfo {author} {\bibfnamefont {A.~R.~P.}\
  \bibnamefont {Lima}}\ and\ \bibinfo {author} {\bibfnamefont {A.}~\bibnamefont
  {Pelster}},\ }\href {\doibase 10.1103/PhysRevA.81.063629} {\bibfield
  {journal} {\bibinfo  {journal} {Phys. Rev. A}\ }\textbf {\bibinfo {volume}
  {81}},\ \bibinfo {pages} {063629} (\bibinfo {year} {2010})}\BibitemShut
  {NoStop}%
\bibitem [{\citenamefont {Zinner}\ and\ \citenamefont
  {Bruun}(2011)}]{Zinner2011a}%
  \BibitemOpen
  \bibfield  {author} {\bibinfo {author} {\bibfnamefont {N.}~\bibnamefont
  {Zinner}}\ and\ \bibinfo {author} {\bibfnamefont {G.}~\bibnamefont {Bruun}},\
  }\href@noop {} {\bibfield  {journal} {\bibinfo  {journal} {European Physical
  Journal D. Atomic, Molecular, Optical and Plasma Physics}\ }\textbf {\bibinfo
  {volume} {65}},\ \bibinfo {pages} {133} (\bibinfo {year} {2011})}\BibitemShut
  {NoStop}%
\bibitem [{\citenamefont {Filinov}\ \emph {et~al.}(2010)\citenamefont
  {Filinov}, \citenamefont {Prokof'ev},\ and\ \citenamefont
  {Bonitz}}]{Filinov2010a}%
  \BibitemOpen
  \bibfield  {author} {\bibinfo {author} {\bibfnamefont {A.}~\bibnamefont
  {Filinov}}, \bibinfo {author} {\bibfnamefont {N.~V.}\ \bibnamefont
  {Prokof'ev}}, \ and\ \bibinfo {author} {\bibfnamefont {M.}~\bibnamefont
  {Bonitz}},\ }\href {\doibase 10.1103/PhysRevLett.105.070401} {\bibfield
  {journal} {\bibinfo  {journal} {Phys. Rev. Lett.}\ }\textbf {\bibinfo
  {volume} {105}},\ \bibinfo {pages} {070401} (\bibinfo {year}
  {2010})}\BibitemShut {NoStop}%
\bibitem [{\citenamefont {Cinti}\ \emph {et~al.}(2010)\citenamefont {Cinti},
  \citenamefont {Jain}, \citenamefont {Boninsegni}, \citenamefont {Micheli},
  \citenamefont {Zoller},\ and\ \citenamefont {Pupillo}}]{Cinti2010a}%
  \BibitemOpen
  \bibfield  {author} {\bibinfo {author} {\bibfnamefont {F.}~\bibnamefont
  {Cinti}}, \bibinfo {author} {\bibfnamefont {P.}~\bibnamefont {Jain}},
  \bibinfo {author} {\bibfnamefont {M.}~\bibnamefont {Boninsegni}}, \bibinfo
  {author} {\bibfnamefont {A.}~\bibnamefont {Micheli}}, \bibinfo {author}
  {\bibfnamefont {P.}~\bibnamefont {Zoller}}, \ and\ \bibinfo {author}
  {\bibfnamefont {G.}~\bibnamefont {Pupillo}},\ }\href {\doibase
  10.1103/PhysRevLett.105.135301} {\bibfield  {journal} {\bibinfo  {journal}
  {Phys. Rev. Lett.}\ }\textbf {\bibinfo {volume} {105}},\ \bibinfo {pages}
  {135301} (\bibinfo {year} {2010})}\BibitemShut {NoStop}%
\bibitem [{\citenamefont {Henkel}\ \emph {et~al.}(2012)\citenamefont {Henkel},
  \citenamefont {Cinti}, \citenamefont {Jain}, \citenamefont {Pupillo},\ and\
  \citenamefont {Pohl}}]{Henkel2012a}%
  \BibitemOpen
  \bibfield  {author} {\bibinfo {author} {\bibfnamefont {N.}~\bibnamefont
  {Henkel}}, \bibinfo {author} {\bibfnamefont {F.}~\bibnamefont {Cinti}},
  \bibinfo {author} {\bibfnamefont {P.}~\bibnamefont {Jain}}, \bibinfo {author}
  {\bibfnamefont {G.}~\bibnamefont {Pupillo}}, \ and\ \bibinfo {author}
  {\bibfnamefont {T.}~\bibnamefont {Pohl}},\ }\href {\doibase
  10.1103/PhysRevLett.108.265301} {\bibfield  {journal} {\bibinfo  {journal}
  {Phys. Rev. Lett.}\ }\textbf {\bibinfo {volume} {108}},\ \bibinfo {pages}
  {265301} (\bibinfo {year} {2012})}\BibitemShut {NoStop}%
\bibitem [{\citenamefont {Babadi}\ and\ \citenamefont
  {Demler}(2011)}]{Babadi2011a}%
  \BibitemOpen
  \bibfield  {author} {\bibinfo {author} {\bibfnamefont {M.}~\bibnamefont
  {Babadi}}\ and\ \bibinfo {author} {\bibfnamefont {E.}~\bibnamefont
  {Demler}},\ }\href {\doibase 10.1103/PhysRevB.84.235124} {\bibfield
  {journal} {\bibinfo  {journal} {Phys. Rev. B}\ }\textbf {\bibinfo {volume}
  {84}},\ \bibinfo {pages} {235124} (\bibinfo {year} {2011})}\BibitemShut
  {NoStop}%
\bibitem [{\citenamefont {Sieberer}\ and\ \citenamefont
  {Baranov}(2011)}]{Baranov2011a}%
  \BibitemOpen
  \bibfield  {author} {\bibinfo {author} {\bibfnamefont {L.~M.}\ \bibnamefont
  {Sieberer}}\ and\ \bibinfo {author} {\bibfnamefont {M.~A.}\ \bibnamefont
  {Baranov}},\ }\href {\doibase 10.1103/PhysRevA.84.063633} {\bibfield
  {journal} {\bibinfo  {journal} {Phys. Rev. A}\ }\textbf {\bibinfo {volume}
  {84}},\ \bibinfo {pages} {063633} (\bibinfo {year} {2011})}\BibitemShut
  {NoStop}%
\bibitem [{\citenamefont {Zinner}\ \emph {et~al.}(2012)\citenamefont {Zinner},
  \citenamefont {Wunsch}, \citenamefont {Pekker},\ and\ \citenamefont
  {Wang}}]{Zinner2012a}%
  \BibitemOpen
  \bibfield  {author} {\bibinfo {author} {\bibfnamefont {N.~T.}\ \bibnamefont
  {Zinner}}, \bibinfo {author} {\bibfnamefont {B.}~\bibnamefont {Wunsch}},
  \bibinfo {author} {\bibfnamefont {D.}~\bibnamefont {Pekker}}, \ and\ \bibinfo
  {author} {\bibfnamefont {D.-W.}\ \bibnamefont {Wang}},\ }\href {\doibase
  10.1103/PhysRevA.85.013603} {\bibfield  {journal} {\bibinfo  {journal} {Phys.
  Rev. A}\ }\textbf {\bibinfo {volume} {85}},\ \bibinfo {pages} {013603}
  (\bibinfo {year} {2012})}\BibitemShut {NoStop}%
\bibitem [{\citenamefont {Parish}\ and\ \citenamefont
  {Marchetti}(2012)}]{Parish2012a}%
  \BibitemOpen
  \bibfield  {author} {\bibinfo {author} {\bibfnamefont {M.~M.}\ \bibnamefont
  {Parish}}\ and\ \bibinfo {author} {\bibfnamefont {F.~M.}\ \bibnamefont
  {Marchetti}},\ }\href {\doibase 10.1103/PhysRevLett.108.145304} {\bibfield
  {journal} {\bibinfo  {journal} {Phys. Rev. Lett.}\ }\textbf {\bibinfo
  {volume} {108}},\ \bibinfo {pages} {145304} (\bibinfo {year}
  {2012})}\BibitemShut {NoStop}%
\bibitem [{\citenamefont {Chan}\ \emph {et~al.}(2010)\citenamefont {Chan},
  \citenamefont {Wu}, \citenamefont {Lee},\ and\ \citenamefont
  {Das~Sarma}}]{Chan2010a}%
  \BibitemOpen
  \bibfield  {author} {\bibinfo {author} {\bibfnamefont {C.-K.}\ \bibnamefont
  {Chan}}, \bibinfo {author} {\bibfnamefont {C.}~\bibnamefont {Wu}}, \bibinfo
  {author} {\bibfnamefont {W.-C.}\ \bibnamefont {Lee}}, \ and\ \bibinfo
  {author} {\bibfnamefont {S.}~\bibnamefont {Das~Sarma}},\ }\href {\doibase
  10.1103/PhysRevA.81.023602} {\bibfield  {journal} {\bibinfo  {journal} {Phys.
  Rev. A}\ }\textbf {\bibinfo {volume} {81}},\ \bibinfo {pages} {023602}
  (\bibinfo {year} {2010})}\BibitemShut {NoStop}%
\bibitem [{\citenamefont {Kestner}\ and\ \citenamefont
  {Das~Sarma}(2010)}]{Kestner2010a}%
  \BibitemOpen
  \bibfield  {author} {\bibinfo {author} {\bibfnamefont {J.~P.}\ \bibnamefont
  {Kestner}}\ and\ \bibinfo {author} {\bibfnamefont {S.}~\bibnamefont
  {Das~Sarma}},\ }\href {\doibase 10.1103/PhysRevA.82.033608} {\bibfield
  {journal} {\bibinfo  {journal} {Phys. Rev. A}\ }\textbf {\bibinfo {volume}
  {82}},\ \bibinfo {pages} {033608} (\bibinfo {year} {2010})}\BibitemShut
  {NoStop}%
\bibitem [{\citenamefont {Ronen}\ and\ \citenamefont
  {Bohn}(2010)}]{Ronen2010a}%
  \BibitemOpen
  \bibfield  {author} {\bibinfo {author} {\bibfnamefont {S.}~\bibnamefont
  {Ronen}}\ and\ \bibinfo {author} {\bibfnamefont {J.~L.}\ \bibnamefont
  {Bohn}},\ }\href {\doibase 10.1103/PhysRevA.81.033601} {\bibfield  {journal}
  {\bibinfo  {journal} {Phys. Rev. A}\ }\textbf {\bibinfo {volume} {81}},\
  \bibinfo {pages} {033601} (\bibinfo {year} {2010})}\BibitemShut {NoStop}%
\bibitem [{\citenamefont {Zhao}\ \emph {et~al.}(2010)\citenamefont {Zhao},
  \citenamefont {Jiang}, \citenamefont {Liu}, \citenamefont {Liu},
  \citenamefont {Zou},\ and\ \citenamefont {Pu}}]{Cheng2010a}%
  \BibitemOpen
  \bibfield  {author} {\bibinfo {author} {\bibfnamefont {C.}~\bibnamefont
  {Zhao}}, \bibinfo {author} {\bibfnamefont {L.}~\bibnamefont {Jiang}},
  \bibinfo {author} {\bibfnamefont {X.}~\bibnamefont {Liu}}, \bibinfo {author}
  {\bibfnamefont {W.~M.}\ \bibnamefont {Liu}}, \bibinfo {author} {\bibfnamefont
  {X.}~\bibnamefont {Zou}}, \ and\ \bibinfo {author} {\bibfnamefont
  {H.}~\bibnamefont {Pu}},\ }\href {\doibase 10.1103/PhysRevA.81.063642}
  {\bibfield  {journal} {\bibinfo  {journal} {Phys. Rev. A}\ }\textbf {\bibinfo
  {volume} {81}},\ \bibinfo {pages} {063642} (\bibinfo {year}
  {2010})}\BibitemShut {NoStop}%
\bibitem [{\citenamefont {Shi}\ \emph {et~al.}(2010)\citenamefont {Shi},
  \citenamefont {Zhang}, \citenamefont {Sun},\ and\ \citenamefont
  {Yi}}]{Shi2010a}%
  \BibitemOpen
  \bibfield  {author} {\bibinfo {author} {\bibfnamefont {T.}~\bibnamefont
  {Shi}}, \bibinfo {author} {\bibfnamefont {J.-N.}\ \bibnamefont {Zhang}},
  \bibinfo {author} {\bibfnamefont {C.-P.}\ \bibnamefont {Sun}}, \ and\
  \bibinfo {author} {\bibfnamefont {S.}~\bibnamefont {Yi}},\ }\href {\doibase
  10.1103/PhysRevA.82.033623} {\bibfield  {journal} {\bibinfo  {journal} {Phys.
  Rev. A}\ }\textbf {\bibinfo {volume} {82}},\ \bibinfo {pages} {033623}
  (\bibinfo {year} {2010})}\BibitemShut {NoStop}%
\bibitem [{\citenamefont {Liao}\ and\ \citenamefont {Brand}(2010)}]{Liao2010a}%
  \BibitemOpen
  \bibfield  {author} {\bibinfo {author} {\bibfnamefont {R.}~\bibnamefont
  {Liao}}\ and\ \bibinfo {author} {\bibfnamefont {J.}~\bibnamefont {Brand}},\
  }\href {\doibase 10.1103/PhysRevA.82.063624} {\bibfield  {journal} {\bibinfo
  {journal} {Phys. Rev. A}\ }\textbf {\bibinfo {volume} {82}},\ \bibinfo
  {pages} {063624} (\bibinfo {year} {2010})}\BibitemShut {NoStop}%
\bibitem [{\citenamefont {Giorgini}\ \emph {et~al.}(1996)\citenamefont
  {Giorgini}, \citenamefont {Pitaevskii},\ and\ \citenamefont
  {Stringari}}]{Giorgini1996a}%
  \BibitemOpen
  \bibfield  {author} {\bibinfo {author} {\bibfnamefont {S.}~\bibnamefont
  {Giorgini}}, \bibinfo {author} {\bibfnamefont {L.~P.}\ \bibnamefont
  {Pitaevskii}}, \ and\ \bibinfo {author} {\bibfnamefont {S.}~\bibnamefont
  {Stringari}},\ }\href {\doibase 10.1103/PhysRevA.54.R4633} {\bibfield
  {journal} {\bibinfo  {journal} {Phys. Rev. A}\ }\textbf {\bibinfo {volume}
  {54}},\ \bibinfo {pages} {R4633} (\bibinfo {year} {1996})}\BibitemShut
  {NoStop}%
\bibitem [{\citenamefont {Giorgini}\ \emph
  {et~al.}(1997{\natexlab{a}})\citenamefont {Giorgini}, \citenamefont
  {Pitaevskii},\ and\ \citenamefont {Stringari}}]{Giorgini1997a}%
  \BibitemOpen
  \bibfield  {author} {\bibinfo {author} {\bibfnamefont {S.}~\bibnamefont
  {Giorgini}}, \bibinfo {author} {\bibfnamefont {L.~P.}\ \bibnamefont
  {Pitaevskii}}, \ and\ \bibinfo {author} {\bibfnamefont {S.}~\bibnamefont
  {Stringari}},\ }\href {\doibase 10.1103/PhysRevLett.78.3987} {\bibfield
  {journal} {\bibinfo  {journal} {Phys. Rev. Lett.}\ }\textbf {\bibinfo
  {volume} {78}},\ \bibinfo {pages} {3987} (\bibinfo {year}
  {1997}{\natexlab{a}})}\BibitemShut {NoStop}%
\bibitem [{\citenamefont {Giorgini}\ \emph
  {et~al.}(1997{\natexlab{b}})\citenamefont {Giorgini}, \citenamefont
  {Pitaevskii},\ and\ \citenamefont {Stringari}}]{Giorgini1997b}%
  \BibitemOpen
  \bibfield  {author} {\bibinfo {author} {\bibfnamefont {S.}~\bibnamefont
  {Giorgini}}, \bibinfo {author} {\bibfnamefont {L.}~\bibnamefont
  {Pitaevskii}}, \ and\ \bibinfo {author} {\bibfnamefont {S.}~\bibnamefont
  {Stringari}},\ }\href {http://dx.doi.org/10.1007/BF02396737} {\bibfield
  {journal} {\bibinfo  {journal} {Journal of Low Temperature Physics}\ }\textbf
  {\bibinfo {volume} {109}},\ \bibinfo {pages} {309} (\bibinfo {year}
  {1997}{\natexlab{b}})},\ \bibinfo {note} {10.1007/BF02396737}\BibitemShut
  {NoStop}%
\bibitem [{\citenamefont {Dalfovo}\ \emph {et~al.}(1997)\citenamefont
  {Dalfovo}, \citenamefont {Giorgini}, \citenamefont {Guilleumas},
  \citenamefont {Pitaevskii},\ and\ \citenamefont {Stringari}}]{Dalfovo1997a}%
  \BibitemOpen
  \bibfield  {author} {\bibinfo {author} {\bibfnamefont {F.}~\bibnamefont
  {Dalfovo}}, \bibinfo {author} {\bibfnamefont {S.}~\bibnamefont {Giorgini}},
  \bibinfo {author} {\bibfnamefont {M.}~\bibnamefont {Guilleumas}}, \bibinfo
  {author} {\bibfnamefont {L.}~\bibnamefont {Pitaevskii}}, \ and\ \bibinfo
  {author} {\bibfnamefont {S.}~\bibnamefont {Stringari}},\ }\href {\doibase
  10.1103/PhysRevA.56.3840} {\bibfield  {journal} {\bibinfo  {journal} {Phys.
  Rev. A}\ }\textbf {\bibinfo {volume} {56}},\ \bibinfo {pages} {3840}
  (\bibinfo {year} {1997})}\BibitemShut {NoStop}%
\bibitem [{\citenamefont {Giorgini}\ \emph {et~al.}(2008)\citenamefont
  {Giorgini}, \citenamefont {Pitaevskii},\ and\ \citenamefont
  {Stringari}}]{Giorgini2008a}%
  \BibitemOpen
  \bibfield  {author} {\bibinfo {author} {\bibfnamefont {S.}~\bibnamefont
  {Giorgini}}, \bibinfo {author} {\bibfnamefont {L.~P.}\ \bibnamefont
  {Pitaevskii}}, \ and\ \bibinfo {author} {\bibfnamefont {S.}~\bibnamefont
  {Stringari}},\ }\href {\doibase 10.1103/RevModPhys.80.1215} {\bibfield
  {journal} {\bibinfo  {journal} {Rev. Mod. Phys.}\ }\textbf {\bibinfo {volume}
  {80}},\ \bibinfo {pages} {1215} (\bibinfo {year} {2008})}\BibitemShut
  {NoStop}%
\bibitem [{\citenamefont {Gerbier}\ \emph
  {et~al.}(2004{\natexlab{a}})\citenamefont {Gerbier}, \citenamefont
  {Thywissen}, \citenamefont {Richard}, \citenamefont {Hugbart}, \citenamefont
  {Bouyer},\ and\ \citenamefont {Aspect}}]{Gerbier2004}%
  \BibitemOpen
  \bibfield  {author} {\bibinfo {author} {\bibfnamefont {F.}~\bibnamefont
  {Gerbier}}, \bibinfo {author} {\bibfnamefont {J.~H.}\ \bibnamefont
  {Thywissen}}, \bibinfo {author} {\bibfnamefont {S.}~\bibnamefont {Richard}},
  \bibinfo {author} {\bibfnamefont {M.}~\bibnamefont {Hugbart}}, \bibinfo
  {author} {\bibfnamefont {P.}~\bibnamefont {Bouyer}}, \ and\ \bibinfo {author}
  {\bibfnamefont {A.}~\bibnamefont {Aspect}},\ }\href@noop {} {\bibfield
  {journal} {\bibinfo  {journal} {Phys.~Rev.~Lett.}\ }\textbf {\bibinfo
  {volume} {92}},\ \bibinfo {pages} {030405} (\bibinfo {year}
  {2004}{\natexlab{a}})}\BibitemShut {NoStop}%
\bibitem [{\citenamefont {Gerbier}\ \emph
  {et~al.}(2004{\natexlab{b}})\citenamefont {Gerbier}, \citenamefont
  {Thywissen}, \citenamefont {Richard}, \citenamefont {Hugbart}, \citenamefont
  {Bouyer},\ and\ \citenamefont {Aspect}}]{Gerbier2004b}%
  \BibitemOpen
  \bibfield  {author} {\bibinfo {author} {\bibfnamefont {F.}~\bibnamefont
  {Gerbier}}, \bibinfo {author} {\bibfnamefont {J.~H.}\ \bibnamefont
  {Thywissen}}, \bibinfo {author} {\bibfnamefont {S.}~\bibnamefont {Richard}},
  \bibinfo {author} {\bibfnamefont {M.}~\bibnamefont {Hugbart}}, \bibinfo
  {author} {\bibfnamefont {P.}~\bibnamefont {Bouyer}}, \ and\ \bibinfo {author}
  {\bibfnamefont {A.}~\bibnamefont {Aspect}},\ }\href@noop {} {\bibfield
  {journal} {\bibinfo  {journal} {Phys. Rev. A}\ }\textbf {\bibinfo {volume}
  {70}},\ \bibinfo {pages} {013607} (\bibinfo {year}
  {2004}{\natexlab{b}})}\BibitemShut {NoStop}%
\bibitem [{\citenamefont {Blaizot}\ and\ \citenamefont
  {Ripka}(1986)}]{BlaizotRipka}%
  \BibitemOpen
  \bibfield  {author} {\bibinfo {author} {\bibfnamefont {J.}~\bibnamefont
  {Blaizot}}\ and\ \bibinfo {author} {\bibfnamefont {G.}~\bibnamefont
  {Ripka}},\ }\href@noop {} {\emph {\bibinfo {title} {{Quantum Theory of Finite
  Systems}}}},\ \bibinfo {edition} {1st}\ ed.\ (\bibinfo  {publisher} {MIT
  Press},\ \bibinfo {address} {Cambridge, Massachusetts},\ \bibinfo {year}
  {1986})\BibitemShut {NoStop}%
\bibitem [{\citenamefont {Sogo}\ \emph {et~al.}(2009)\citenamefont {Sogo},
  \citenamefont {He}, \citenamefont {Miyakawa}, \citenamefont {Yi},
  \citenamefont {Lu},\ and\ \citenamefont {Pu}}]{Sogo2009a}%
  \BibitemOpen
  \bibfield  {author} {\bibinfo {author} {\bibfnamefont {T.}~\bibnamefont
  {Sogo}}, \bibinfo {author} {\bibfnamefont {L.}~\bibnamefont {He}}, \bibinfo
  {author} {\bibfnamefont {T.}~\bibnamefont {Miyakawa}}, \bibinfo {author}
  {\bibfnamefont {S.}~\bibnamefont {Yi}}, \bibinfo {author} {\bibfnamefont
  {H.}~\bibnamefont {Lu}}, \ and\ \bibinfo {author} {\bibfnamefont
  {H.}~\bibnamefont {Pu}},\ }\href
  {http://stacks.iop.org/1367-2630/11/i=5/a=055017} {\bibfield  {journal}
  {\bibinfo  {journal} {New Journal of Physics}\ }\textbf {\bibinfo {volume}
  {11}},\ \bibinfo {pages} {055017} (\bibinfo {year} {2009})}\BibitemShut
  {NoStop}%
\bibitem [{\citenamefont {Ronen}\ \emph
  {et~al.}(2006{\natexlab{b}})\citenamefont {Ronen}, \citenamefont
  {Bortolotti},\ and\ \citenamefont {Bohn}}]{Ronen2006a}%
  \BibitemOpen
  \bibfield  {author} {\bibinfo {author} {\bibfnamefont {S.}~\bibnamefont
  {Ronen}}, \bibinfo {author} {\bibfnamefont {D.~C.~E.}\ \bibnamefont
  {Bortolotti}}, \ and\ \bibinfo {author} {\bibfnamefont {J.~L.}\ \bibnamefont
  {Bohn}},\ }\href@noop {} {\bibfield  {journal} {\bibinfo  {journal} {Phys.
  Rev. A}\ }\textbf {\bibinfo {volume} {74}},\ \bibinfo {eid} {013623}
  (\bibinfo {year} {2006}{\natexlab{b}})}\BibitemShut {NoStop}%
\bibitem [{\citenamefont {Bisset}\ \emph {et~al.}(2011)\citenamefont {Bisset},
  \citenamefont {Baillie},\ and\ \citenamefont {Blakie}}]{Bisset2011}%
  \BibitemOpen
  \bibfield  {author} {\bibinfo {author} {\bibfnamefont {R.~N.}\ \bibnamefont
  {Bisset}}, \bibinfo {author} {\bibfnamefont {D.}~\bibnamefont {Baillie}}, \
  and\ \bibinfo {author} {\bibfnamefont {P.~B.}\ \bibnamefont {Blakie}},\
  }\href {\doibase 10.1103/PhysRevA.83.061602} {\bibfield  {journal} {\bibinfo
  {journal} {Phys. Rev. A}\ }\textbf {\bibinfo {volume} {83}},\ \bibinfo
  {pages} {061602(R)} (\bibinfo {year} {2011})}\BibitemShut {NoStop}%
\bibitem [{\citenamefont {Donner}\ \emph {et~al.}(2007)\citenamefont {Donner},
  \citenamefont {Ritter}, \citenamefont {Bourdel}, \citenamefont {{\"O}ttl},
  \citenamefont {K{\"o}hl},\ and\ \citenamefont {Esslinger}}]{Donner2007a}%
  \BibitemOpen
  \bibfield  {author} {\bibinfo {author} {\bibfnamefont {T.}~\bibnamefont
  {Donner}}, \bibinfo {author} {\bibfnamefont {S.}~\bibnamefont {Ritter}},
  \bibinfo {author} {\bibfnamefont {T.}~\bibnamefont {Bourdel}}, \bibinfo
  {author} {\bibfnamefont {A.}~\bibnamefont {{\"O}ttl}}, \bibinfo {author}
  {\bibfnamefont {M.}~\bibnamefont {K{\"o}hl}}, \ and\ \bibinfo {author}
  {\bibfnamefont {T.}~\bibnamefont {Esslinger}},\ }\href {\doibase
  10.1126/science.1138807} {\bibfield  {journal} {\bibinfo  {journal}
  {Science}\ }\textbf {\bibinfo {volume} {315}},\ \bibinfo {pages} {1556}
  (\bibinfo {year} {2007})}\BibitemShut {NoStop}%
\end{thebibliography}
\end{document}